\documentclass[%
 aip,
 jmp,
 amsmath,amssymb,
preprint,%
]{revtex4-1}

\usepackage{graphicx}
\usepackage{dcolumn}
\usepackage{bm}
\usepackage{amsmath}
\usepackage{mathtools} 

\usepackage[utf8]{inputenc}
\usepackage[T1]{fontenc}
\usepackage{mathptmx}
\usepackage{chngcntr}
\usepackage{etoolbox}
\usepackage{blindtext}
\usepackage[colorlinks=true,linkcolor=black,citecolor=black,urlcolor=black]{hyperref}
\makeatother
\begin{document}
\preprint{AIP/123-QED}

\title{Numerical investigation of kinetic instabilities in BGK equilibria under collisional effects}

\author{S. Zanelli}
\email{sofia.zanelli@unical.it}
\affiliation{Dipartimento di Fisica, Universitá della Calabria, Rende, I-87036, Italy}
\author{G. Celebre}
\affiliation{Dipartimento di Fisica, Universitá della Calabria, Rende, I-87036, Italy}
\author{S. Servidio}
\affiliation{Dipartimento di Fisica, Universitá della Calabria, Rende, I-87036, Italy}
\author{F. Valentini}
\affiliation{Dipartimento di Fisica, Universitá della Calabria, Rende, I-87036, Italy}

\date{\today}

\begin{abstract}
\textbf{ABSTRACT}\\

An unstable one-dimensional Bernstein–Greene–Kruskal (BGK) mode has been studied through high-precision numerical simulations. The initial turbulent, periodic equilibrium state is obtained by solving a Vlasov-Poisson system for initially thermalized electrons, with the addition of an external electric field able to trigger undamped, high-amplitude electron acoustic waves (EAWs). Once the external field is turned off, resonant particles are trapped in a stationary two-hole phase-space configuration. This equilibrium scenario is perturbed by some large-scale density noise, leading to an electrostatic instability with the merging of vortices into a final one-hole state. Numerical runs investigate several features of this regime, focusing on the dependence of the instability trigger time and growth rate on the rate of short-range collisions and grid resolution. According to Landau theory for weakly inhomogeneous equilibria, we observe that the growth rate of the instability depends only on the slope of the distribution function in the resonant region. Conversely, the onset time of the instability is affected by the collisional rate, which is able to postpone the onset of the instability. Moreover, by extending the simulations to a long-time scale, we investigate the saturation stage of the instability, which can be analyzed through the Hermite spectral analysis. In collisionless simulations where grid effects are negligible, the Hermite spectrum follows a power law typical of a constant enstrophy flux scenario. Otherwise, if collisional effects become significant, a cutoff is observed at high Hermite modes, leading to a decaying trend.

\end{abstract}

\maketitle

\section{\label{sec:level1} Introduction}
The dynamics of collisionless plasmas is highly conditioned by turbulence, which plays a fundamental role in governing energy transfer and dissipation. Turbulence can be described, in a kinetic framework, through a nonlinear cross-scale coupling associated with a cascade of free energy, or enstrophy, in the phase space of the distribution function of charged particles \cite{schekochihin2016phase,servidio2017magnetospheric,celebre2023phase,nastac2024phase,nastac2025universal}. However, how turbulence is generated and what role it plays in energy transport remain open questions \cite{bruno2013solar}. In order to investigate such dynamics, a common and effective way is to find and analyze the nonlinear solutions of the Vlasov-Poisson system of differential equations, which describes the complex kinetic dynamics found in laboratory and space plasmas. The system is valid only in an electrostatic regime, where magnetic fluctuations can be neglected because the plasma beta — defined as the ratio between plasma pressure and magnetic pressure — is very low \cite{krall1974principles}.

In this framework, in 1957, Bernstein, Greene and Kruskal \cite{bernstein1957exact} showed the existence of stationary solutions depending only on single-particle energy. These solutions, referred to as BGK modes, characterize a large class of equilibrium states in which particles play an important role in shaping the structure of the electric field. The most important mechanism underlying BGK modes is the self-consistent equilibrium between a localized electrostatic potential and the particle density, sustained by a continuous distribution of kinetic energy \cite{krall1974principles}. In this scenario, particles can be trapped inside potential wells, and the nonlinearity of the stationary solution is realized through a distortion of the particle distribution function (PDF) in the vicinity of the wave phase velocity \cite{o1965collisionless}. In phase space, this distortion appears as a localized plasma region characterized by a density lower than that of the surrounding plasma, caused by the reduced phase-space density of trapped electrons. This region is known as an electron hole \cite{hutchinson2017electron}. In collisionless plasmas, these structures lead to the formation of a plateau in the resonant region \cite{kennel1966velocity}, with the suppression of Landau damping, resulting in scenarios with long-lived, undamped electrostatic oscillations. By appropriately selecting the distribution of trapped particles, self-consistent equilibrium states can be constructed for essentially arbitrary potential profiles \cite{turikov1984electron, muschietti1999analysis}.

Electron holes are deeply studied because of their fundamental role in plasma dynamics, being involved, for instance, in the phenomenology of magnetic reconnection, shocks, and plasma wakes. Despite their small spatial scale—typically only a few Debye lengths—and their fast oscillations over time, electron holes can strongly influence the macroscopic evolution of plasma systems \cite{hutchinson2017electron}. These structures have been extensively investigated both in laboratory and space plasmas. Laboratory experiments, like those by Saeki et al. \cite{saeki1979formation}, have shown their formation and coalescence in magnetized, collisionless plasmas and identified them as BGK-type equilibria sustained by trapped electrons. Space observations with the FAST satellite \cite{ergun1998fast} have also reported large-amplitude solitary structures identified as electron holes in the auroral zone, where they contribute to sustaining parallel electric fields and modulating energetic electron fluxes. By appropriately selecting the number and characteristics of trapped particles, self-consistent equilibrium states can be constructed for essentially arbitrary potential profiles \cite{schamel1971stationary, schamel1972non}.

BGK solutions have continued to attract considerable interest, as they may represent the final, saturated state of instabilities that are stabilized by particle trapping within the potential wells generated by growing waves \cite{shoucri1979nonlinear, hannibal1994bifurcation}. Over the years, stationary BGK modes have been studied from various theoretical perspectives. In 2000, Manfredi et al. \cite{manfredi2000stability} analyzed the stability of BGK modes in the regime of small electric potential. Their work revealed that even single-hole BGK structures can become unstable, in contrast with previous studies that had only reported instability in multi-hole structures \cite{berk1970phase, ghizzo1988stability}. In another study, Valentini \cite{valentini2008damping} analyzed the impact of Coulomb collisions on the stability of BGK modes. Revisiting the earlier work by Zakharov and Karpman \cite{zakharov1963nonlinear}, he demonstrated that collisions tend to destroy the BGK structure by driving the system back toward a Maxwellian equilibrium. This emphasizes the important role of BGK modes in studying the nonlinear and turbulent dynamics of space plasmas.

Another interesting area of research has focused on the dynamics of electron acoustic waves (EAWs). These nonlinear plasma effects were first reported by Holloway and Dorning \cite{holloway1991undamped} in 1991, and are characterized by the effect of electron trapping, which inhibits the linear Landau damping \cite{landau196561}. In 2006, Valentini et al. \cite{valentini2006excitation} investigated the excitation and stability of the nonlinear electrostatic branch using particle-in-cell (PIC) simulations. Their analysis focused on the long-term phase-space dynamics arising from the resonant interaction between EAWs and kinetic electrons, enabled by the presence of an external electric field that introduces nonlinearity into the system. More recently, Valentini et al. \cite{valentini2025decay} further explored the excitation and decay mechanisms of this electrostatic branch, with particular attention to the initial formation of the waves, the onset of decay instabilities, and the role of vortex merging in phase space. Their findings show that EAWs can be excited by a small amplitude electric field, which triggers a resonant interaction with the electrons. This represents the electric counterpart of waves associated with ion trapping investigated in Ref.~\onlinecite{valentini2011new}. This interaction creates a diffusive process in velocity space that finally ends up with the creation of a single coherent hole in the electron distribution function. In the long-time limit, the system evolves toward a stationary nonlinear state, which can be identified as a BGK mode. Under small perturbations, this structure becomes unstable, leading to a vortex merging driven by phase-space transport.

EAWs are also usually found in space and laboratory plasmas. Lalti et al. \cite{lalti2025debye}, for example, described the occurrence of EAW-like electrostatic waves in space plasmas, more specifically along quasi-perpendicular collisionless shocks. These waves, which are usually at Debye-scale wavelengths and frequencies less than the ion plasma frequency, are important to shock microphysics by aiding anomalous dissipation and electron thermalization. In a different setting, Anderegg et al. \cite{anderegg2009wave, anderegg2009electron, affolter2018trapped} demonstrated that EAW-type modes can also be excited in pure ion plasmas confined in laboratory Penning traps. In this scenario, the waves appear as standing modes with tunable frequencies, excited by external electrodes. Even though no electrons are present, the modes show a nonlinear kinetic behavior similar to EAWs: particle trapping and redistribution in velocity space allow the plasma to adapt and come into resonance with the driving frequency.

In the present work, we analyze the properties of the unstable BGK state in both collisionless and weakly collisional plasma regimes. We deduce that the growth rate of this instability is (almost) independent of the rate of collisions, showing a parallelism with Landau’s theory for homogeneous equilibria \cite{landau196561}. In contrast, the trigger time for the instability is significantly influenced by collisions, which can delay the vortex merging. Additionally, we analyzed the saturation phase of the instability using Hermite spectral analysis to characterize the nature of the resulting turbulent state \cite{schumer1998vlasov, pezzi2018velocity}. Indeed, this investigation permits to detect the amount of enstrophy (i.e., the fluctuation of the second Casimir invariant with respect to the equilibrium PDF, see Refs.~\onlinecite{servidio2017magnetospheric,celebre2023phase}) stored at each scale.

Some recent works about kinetic plasma turbulence show how the formation of small phase-space structures can be modeled by assuming a fluid-like enstrophy (or, equivalently, free energy) cascade from large to small scales. This topic was first introduced by Schekochihin et al. \cite{schekochihin2016phase}, who theoretically derived enstrophy spectra associated with linear and nonlinear stationary regimes. Servidio et al. \cite{servidio2017magnetospheric} deduced the kinetic turbulence associated with systems where the enstrophy flux is constant between the large (forcing) scales and the small (collisional) scales, where free energy is ultimately dissipated. They also got observational evidence in support of their hypotheses by applying a three-dimensional Hermite transform to data from the Magnetospheric Multiscale (MMS) mission, revealing highly structured velocity distributions. Hermite decomposition is well suited for decomposing the PDF profiles in velocity space, considering that the thermal component of the velocity distribution is concentrated in the lowest Hermite mode, unlike what occurs in a traditional Fast Fourier Transform (FFT), where finer structures populate the higher modes. According to this, we apply the Hermite transform to the PDFs obtained from the simulations presented in this work, using the numerical technique described in Refs.~\onlinecite{servidio2017magnetospheric,celebre2023phase}.

After the nonlinear saturation of the instability, the system remains quasi-stationary and shows well-developed small-scale structures.
Therefore, it represents an ideal turbulent scenario to be analyzed through the Hermite transform in velocity space and to study whether BGK modes respect the aforementioned cascade models as other nonlinear regimes analyzed in previous studies \cite{servidio2017magnetospheric, pezzi2018velocity, pezzi2019fourier, celebre2023phase, nastac2024phase, nastac2025universal}.  
This allows us to observe whether, in an intermediate range of velocity scales, the flux of enstrophy is actually conserved. 
Furthermore, we discuss the evolution of the spectrum on collisional scales by employing the collision operator that efficiently suppresses fine-scale structures and brings the system close to the state of local thermal equilibrium. In this context, particular attention is devoted to the role of enstrophy cascades in regions inside and outside phase-space vortices, in order to explore how large-scale structures may influence turbulent mechanisms and the scaling of the Hermite spectrum \cite{pezzi2018velocity, pezzi2019fourier, nastac2024phase, nastac2025universal}.

This paper is organized as follows: in Sec.~\ref{sec:numericalmodel}, we present the numerical code employed to solve Vlasov-Poisson system for kinetic electrons and motionless ions, together with all the parameters of the simulations. Sec.~\ref{sec:results} is devoted to the presentation of the main results of our simulations, whereas, in Sec.~\ref{sec:summary}, we discuss our conclusions. Additionally, an \hyperref[appendice]{Appendix} is included to describe a theoretical method for predicting the instability growth rate associated with spatially-dependent BGK equilibria.

\section{Numerical setup}\label{sec:numericalmodel}
In this section, we describe the algorithm used to solve the 1D-1V Vlasov-Poisson system, treating electrons as a kinetic species and ions as a fixed background charge. The system, in dimensionless units, is written as follows:
\begin{gather}
\frac{\partial f}{\partial t} + v\frac{\partial f}{\partial x} - E\frac{\partial f}{\partial v} = \frac{\partial f}{\partial t}\Big|_{coll}, \label{eqn:f} \\[5pt]
\frac{\partial E}{\partial x} = 1 - \int_{-\infty}^{+\infty} f \, dv, \label{eqn:E}
\end{gather}
where $f=f\left(x,v,t\right)$ represents the electron distribution function and $E=E\left(x,t\right)$ is the self-consistent electric field. In Eqs.~\eqref{eqn:f}-\eqref{eqn:E}, time is normalized by the inverse electron plasma frequency $\omega_{p,e}^{-1}$, velocities by the electron thermal speed $v_{th,e}$ and lengths by the electron Debye length $\lambda_{D,e}$. Therefore, the electric field is scaled by $\sqrt{4\pi_{0}m_{e}v_{th,e}^{2}}$ and the electron distribution function by $n_{0}/v_{th,e}$, where $n_{0}$ is the equilibrium density and $m_{e}$ is the electron mass. In Eq.~\eqref{eqn:f} the collisional term $\frac{\partial f}{\partial t}\Big|_{coll}$ is modeled by the Dougherty operator \cite{dougherty1964model, dougherty1967model}, which, according to this normalization, takes the form:
\begin{equation}
    \frac{\partial f}{\partial t}\Big|_{coll}= \nu\left(n_{e},T_{e}\right)\frac{\partial }{\partial v}\left[T_{e}\left(f\right)\frac{\partial f}{\partial v}+\left(v-U_{e}\left(f\right)\right)f\right],
\end{equation}
where $n_{e}=\int f \, dv$ is the electron density, $U_{e} = \int vf \, dv/n_{e}$ represents the electron bulk speed and $T_{e}= \int \left( v-U_{e}\right)^{2}f \,dv/n_{e}$ is the electron temperature; in addition, $\nu \left( n_{e}, T_{e}\right)= \nu_{0}n_{e}/T^{3/2}$ is the collision frequency, where $\nu_{0}=-g\ln{g}/(24\pi)$ (g is the plasma parameter). 

The simulation phase space is discretized by $N_{x}=8192$ grid points in the spatial domain and $N_{v}=8001$ grid points in the velocity domain. The velocity domain is bounded by setting the PDF to zero for $|v|>v_{max}$, with $v_{max}=6$. The spatial domain is periodic, with a size set to $L_{x}=40$, which corresponds to a wave number of $k_{0}=2\pi/L_{x} \simeq 0.16$. In all numerical runs, the system evolves until the maximum time $t_{max} = 10500$. The initial condition $f_{eq}(x,v)$ for the equilibrium distribution function in each simulation is a BGK structure obtained as the result of the excitation of an EAW as in Ref.~\onlinecite{valentini2025decay}, consisting of two phase-space vortices propagating in the positive $x$ direction with velocity equal to the phase speed of the excited EAW $v_{\phi}\simeq 1.7$ [Fig.~\ref{fig1} (a)].
The initial equilibrium is perturbed by sinusoidal density noise introduced at $t=0$ that oscillates with wavelength $L_x$, i.e., $f\left(x,v,t=0 \right)= \left[ 1 + A\cos\left(k_{0}x\right) \right] f_{eq}(x,v)$, with $A=10^{-12}$.

In our simulations, we analyze the distribution function $f$ using Hermite decomposition, as described in Refs.~\onlinecite{servidio2017magnetospheric,celebre2023phase}. In particular, given the basis $\lbrace \psi_m\rbrace$, with $\psi_m(v)=H_{m}\left( v\right)e^{-v^{2}/2}/\sqrt{2^{m}m!\sqrt{\pi}}$ and $H_m$ being the $m$-th Hermite polynomial, we define:
\begin{equation}
f_{m}(x,t) = \int_{-\infty}^{+\infty} f(x,v,t) \psi_{m}(v) \, dv.
\end{equation}
In our case, the PDF is defined on a homogeneous grid \((x_j, v_l, t_p)\). Through interpolation, we move from the velocity grid points \(v_l\)'s to a new set of points \(v_q\)'s, which represent the roots of the $M$-th Hermite polynomial $H_{M}(v)$ within $\left[-v_{max}, v_{max}\right]$. The discrete Hermite coefficients, obtained by means of an appropriate quadrature (see also Ref.~\onlinecite{golub1969calculation}), are then given by:
\begin{equation}
f_{m}(x_{j},t_{p})= \sum_{q=1}^{M}w_{q}g_{m}\left(x_{j},v_{q},t_{p} \right),
\end{equation}
where the weights $w_{q}$'s are calculated as $w_{q}=2^{M-1}M!\sqrt{\pi}/(M^{2}H_{M-1}^{2}(v_{q}))$ and $g_{m}(v)$ is defined as
\begin{equation}
    g_{m}\left(v\right)=\frac{f\left(v\right)H_{m}\left( v\right)e^{v^{2}/2}}{\sqrt{2^{m}m!\sqrt{\pi}}},
\end{equation}
which represents the contribution of the distribution function weighted by the Hermite transform \cite{servidio2017magnetospheric,celebre2023phase}.
For each simulation, we set $M=800$. This choice of $M$ and $v_{max}$ allows us to investigate a broad velocity range with a limited numerical error. This is confirmed by preliminary convergence tests on some cases treated in this paper, where these parameters are tuned separately. On one hand, the values of $f_m$ obtained with $v_{max} = 6$ and $v_{max} = 30$ differ only at the level of machine precision, since the PDF is already negligible for $| v | > 6$. On the other hand, Hermite spectra show comparable trends for $M$ between $100$ and $800$, with a reasonable round-off error in Parseval’s identity that is smaller than $10^{-5}$ when $M=800$.

The Vlasov equation is solved, for the collisionless case, through the splitting scheme introduced by Cheng and Knorr \cite{cheng1976integration}, generalized as in Ref.~\onlinecite{filbet2002numerical} in the presence of collisions (see also Refs.~\onlinecite{valentini2005numerical,valentini2007hybrid, pezzi2013eulerian,pezzi2014erratum,pezzi2019fourier,celebre2023phase}). The evolution in both physical and velocity space is evolved through the implementation of an upwind finite-volume scheme that draws from the work by Van Leer \cite{van1997towards}, achieving third-order accuracy \cite{mangeney2002numerical, valentini2011new, perrone2012vlasov, valentini2012undamped, valentini2013response}. The Poisson equation is solved through a 1D FFT, enabling a fast and efficient computation of the electric field. To ensure numerical stability, the time step $\Delta t$ is selected according to the Courant–Friedrichs–Lewy (CFL) stability criterion \cite{courant1967partial}. Throughout all simulations presented in this work, the energy variation consistently remains below $10^{-4}\%$.

\section{Numerical results}\label{sec:results}
\subsection{Phase-space electron distribution functions} \label{subA}
Following and extending Valentini et al. \cite{valentini2025decay}, we describe the time evolution of the unstable one-dimensional BGK mode in both the collisionless and weakly collisional regimes. The first key aspect to examine is the evolution of the electron distribution function in phase space. The application of the large-scale density perturbation induces an electrostatic instability, regulated by the trapping/de-trapping processes of electrons in the potential wells of the waves, as discussed in Ref.~\onlinecite{valentini2025decay}. This sets in a diffusive process in phase space, through which the initial two holes gradually evolve. In the limit of large times, they tend to merge into a single coherent hole around $v_{\phi}=1.7$, the phase velocity of the waves.
\begin{figure*}[b]
   \begin{center}
        \includegraphics[]{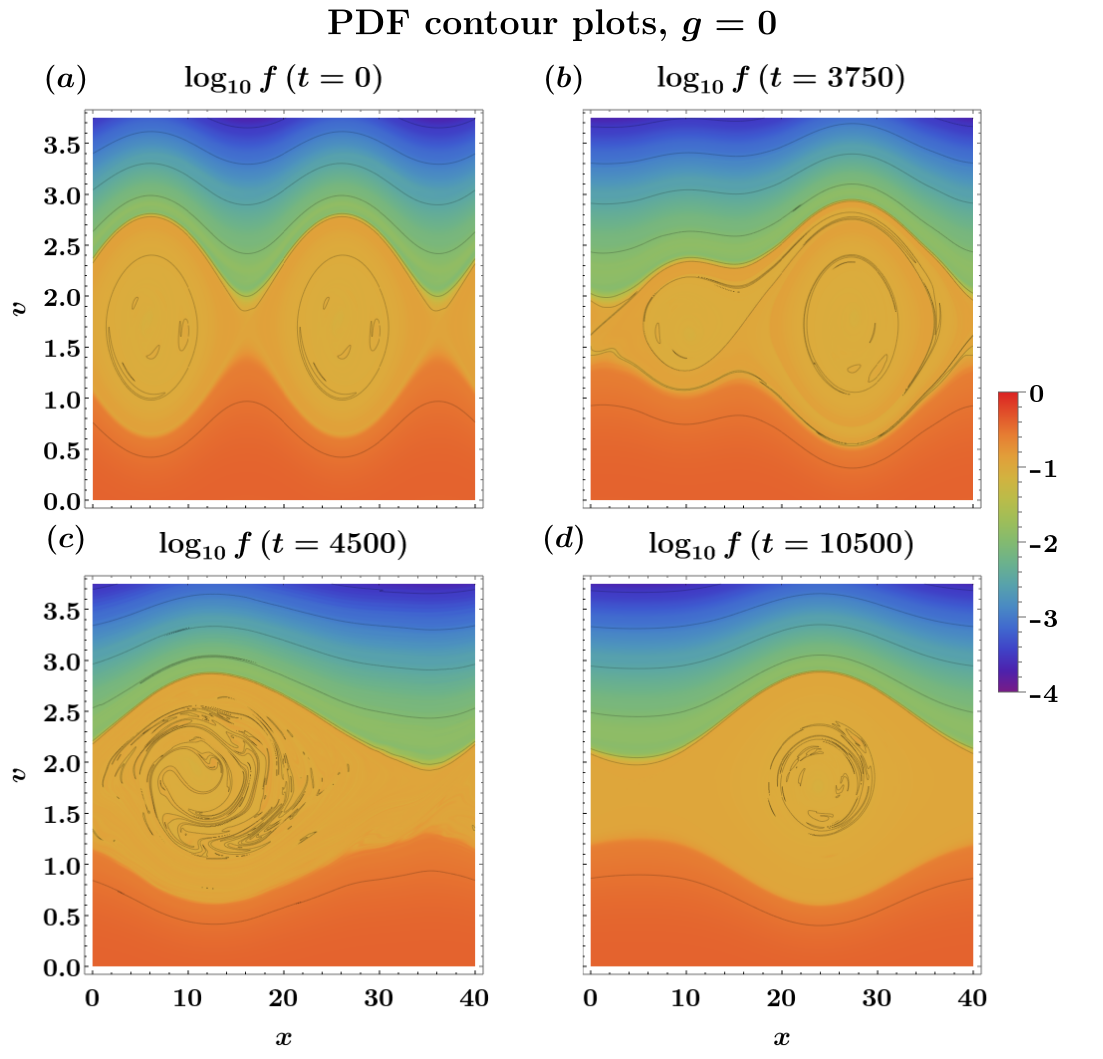}
        \caption{Contour plots of the phase-space electron distribution function at $t=0$ [panel (a)], $t=3750$ [panel (b)], $t=4500$ [panel (c)], and $t=10500$ [panel (d)] for $g=0$ (collisionless case).}
        \label{fig1}
    \end{center}
\end{figure*}
However, the onset of the instability is affected by the presence of collisions. Specifically, collisions tend to delay the development of the instability and, consequently, the merging of the vortices. This behavior can be observed in Figs.~\ref{fig1} and \ref{fig2}, which show $f$ contour plots during the evolution of the system, in the collisionless case ($g=0$) and in the weakly collisional case ($g=10^{-5}$), respectively. 

Focusing on Fig.~\ref{fig1}, at the initial time, two vortical structures can be identified, which, during the merging phase [panel (b)], tend to coalesce and form a single coherent hole [panel (c)] that remains stable until the end of the simulation [panel (d)]. A similar behavior is observed in Fig.~\ref{fig2}. Starting from the same initial condition, we observe that the collisional term indeed delays the onset of the instability, with the merging occurring later than in the collisionless case. While for $g=0$ the merging takes place at $t=3750$, it occurs at $t=4500$ for $g=10^{-5}$.
\begin{figure*}[b!]
   \begin{center}
        \includegraphics[]{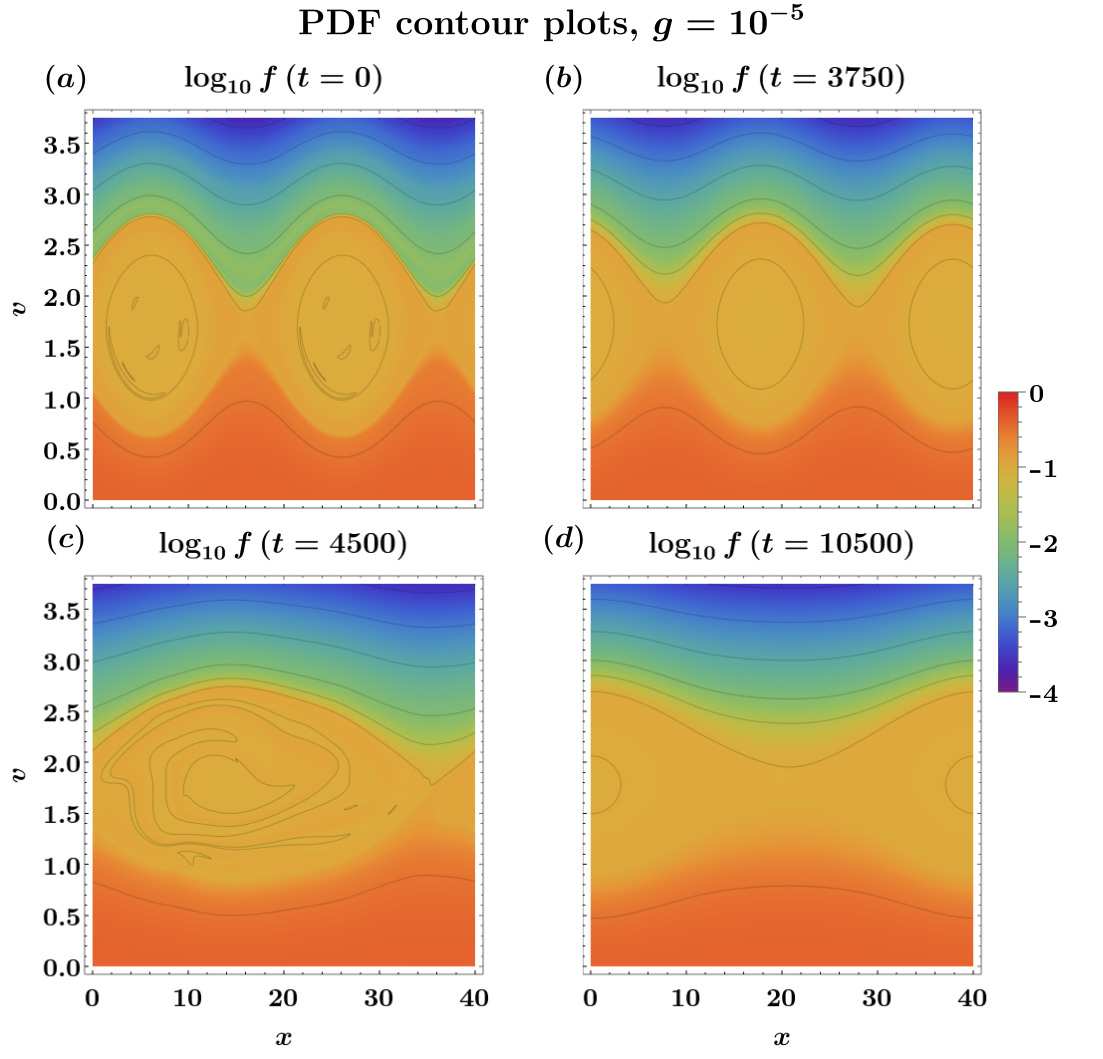}
        \caption{Contour plots of the phase-space electron distribution function at $t=0$ [panel (a)], $t=3750$ [panel (b)], $t=4500$ [panel (c)], and $t=10500$ [panel (d)] for $g=10^{-5}$ (weakly collisional case).}
        \label{fig2}
    \end{center}
\end{figure*} 
This shows that the dynamics of mode $n=1$ is influenced by the presence of collisions. In particular, collisions affect the electron distribution function in phase space by delaying the nonlinear stage. As discussed in Ref.~\onlinecite{valentini2025decay}, the instability is triggered by trapped particles, associated with fine-scale structures in velocity space. Collisions introduce velocity-space diffusion that drives the system toward local thermodynamic equilibrium, thereby smoothing out these gradients. Therefore, the collisional term works to reduce the number of trapped particles, leading to a clear delay compared to the collisionless case, where fine structures persist and the merging develops more rapidly.

\subsection{Growth rate and trigger time in collisionless and weakly collisional regimes}
\begin{figure*}[b!]
   \begin{center}
        \includegraphics[]{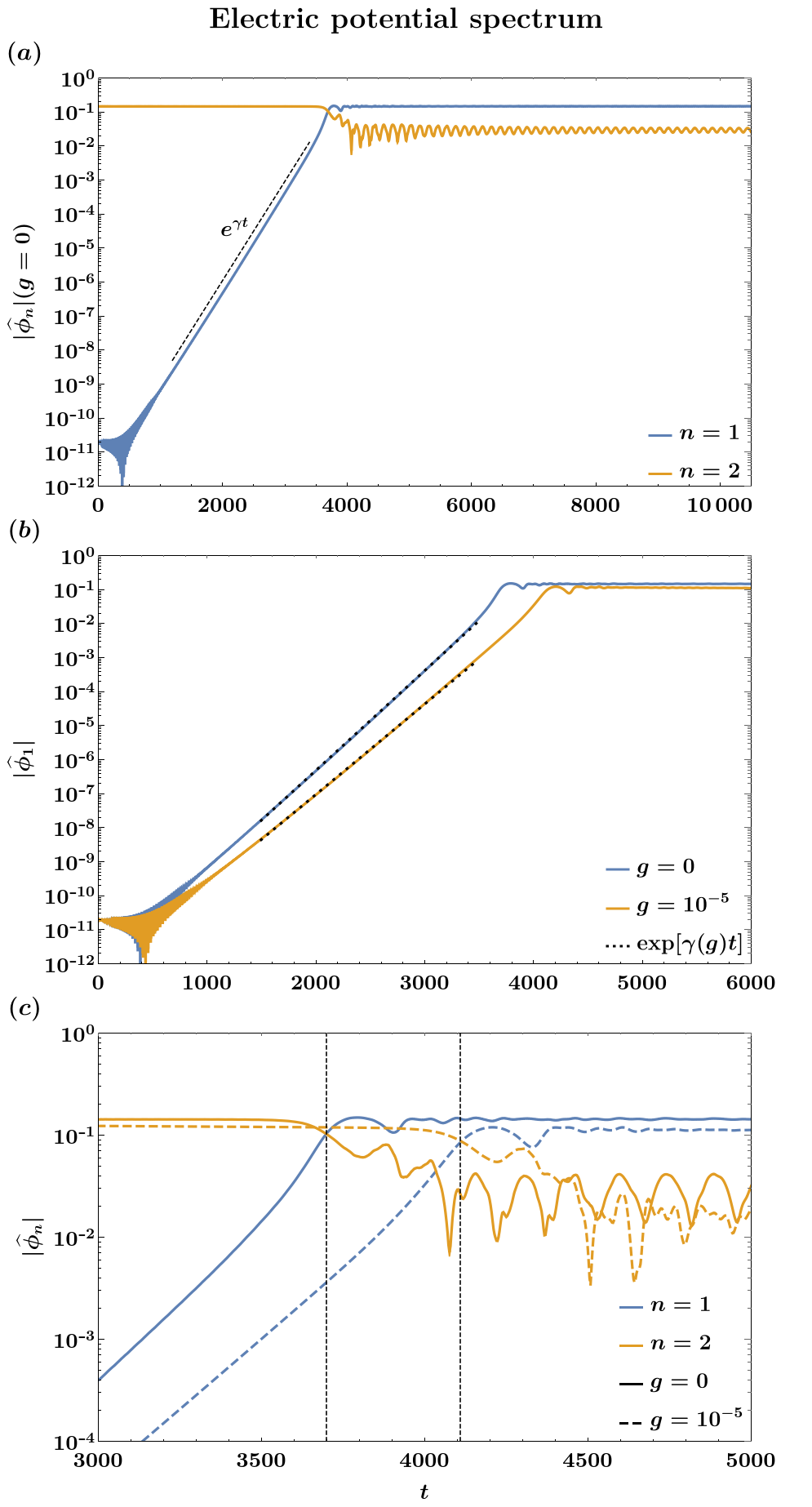}
        \caption{Time evolution of the amplitude of the electric potential Fourier components $|\widehat{\phi}_1|$ and $|\widehat{\phi}_2|$ in the collisional and collisionless cases. Panel (a): mode 1 (blue curve) and mode 2 (orange curve) for $g=0$, compared with the typical trend $e^{\gamma t}$ (black dashed line). Panel (b): $| \widehat{\phi}_1 |$ evaluated for $g=0$ (blue curve) and $g=10^{-5}$ (orange curve), compared with the best fits $e^{\gamma t}$ (black dotted lines). Panel (c): comparison between $| \widehat{\phi}_1 |$ (blue lines) and $| \widehat{\phi}_2|$ (orange lines) in the merging phase for for $g=0$ (solid lines) and $g=10^{-5}$ (dashed lines). The crossing time $\tau_{cross}$ is highlighted by black dashed lines.}
        \label{fig3}
    \end{center}
\end{figure*}

The delay seen in the electron PDF caused by the action of collisional effects can be further confirmed by analyzing the time development of the electric potential spectrum $\widehat{\phi}_n$ for wavevectors $k_n=nk_0$ with $n = 1, 2$ and $k_{0}=2\pi/L_{x}$, plotted in Fig.~\ref{fig3} for both the collisionless and weakly collisional case. Panel (a) refers to the $g=0$ case. Fourier mode $n=1$, after the instability has been triggered, grows exponentially in time according to the law $e^{\gamma t}$ with $\gamma \simeq 6.74 \times 10^{-3}$. The two modes cross at time $t \simeq 3700$, corresponding to the onset of the merging phase. The same trend is observed for $g=10^{-5}$ where mode $n=1$ grows with $\gamma \simeq 6.11 \times 10^{-3}$. In this case, the crossing of the two modes occurs at $t \simeq 4110$. From the comparison between the collisionless and collisional cases in Fig.~\ref{fig3} (b), it is evident that the instability growth rate is not significantly affected by the inclusion of collisions. Indeed, the growth rate associated with the Fourier mode $n = 1$ varies by only about $9\%$, decreasing from $ \simeq 6.74 \times 10^{-3}$ to $\simeq 6.11 \times 10^{-3}$.

This supports the conclusion that collisions do not introduce substantial changes in the instability growth rate. Considering that the duration of the transient phase before the exponential growth does not seem to change with $g$, the variation of $\gamma$ is reflected on the delay of the crossing time $\tau_{cross}$, i.e., the time at which the Fourier amplitudes of modes $n=1$ and $n=2$ intersect. Indeed, it increases by approximately the same percentage, $11\%$, from $\simeq 3700$ to $\simeq 4110$ [panel (c)]. This also provides further evidence of the delay seen in the vortex merging of the electron distribution function given in Sec.~\ref{subA}.
The dependence of $\tau_{cross}$ on $g$ could be affected by the suppression of the mode $n=2$ due to the presence of collisions. Indeed, we would expect that it tends to dissipate more rapidly than in the collisionless case, lowering the threshold value that mode 1 must reach to become dominant. If mode 1 grows exponentially after a transient $\tau_0$ before it intersects with mode 2 at the value $\phi_{cross}=|\widehat{\phi}_{1}(\tau_{cross})|$, then $\tau_{cross} \sim \ln(\phi_{cross} / |\widehat{\phi}_{1}(\tau_{0})|) / \gamma+\tau_{0}$. Observing the two modes in the collisionless ($g=0$) and collisional ($g=10^{-5}$) cases in the aforementioned plot, we notice how $\phi_{cross}$ decreases by only $15\%$ with $g$. From the previous formula, we can deduce that the impact of this variation on $\tau_{cross}$ is about $0.6\%$, which is much smaller than the variation driven by $\gamma$.

\begin{figure*}[b!]
   \begin{center}
        \includegraphics[]{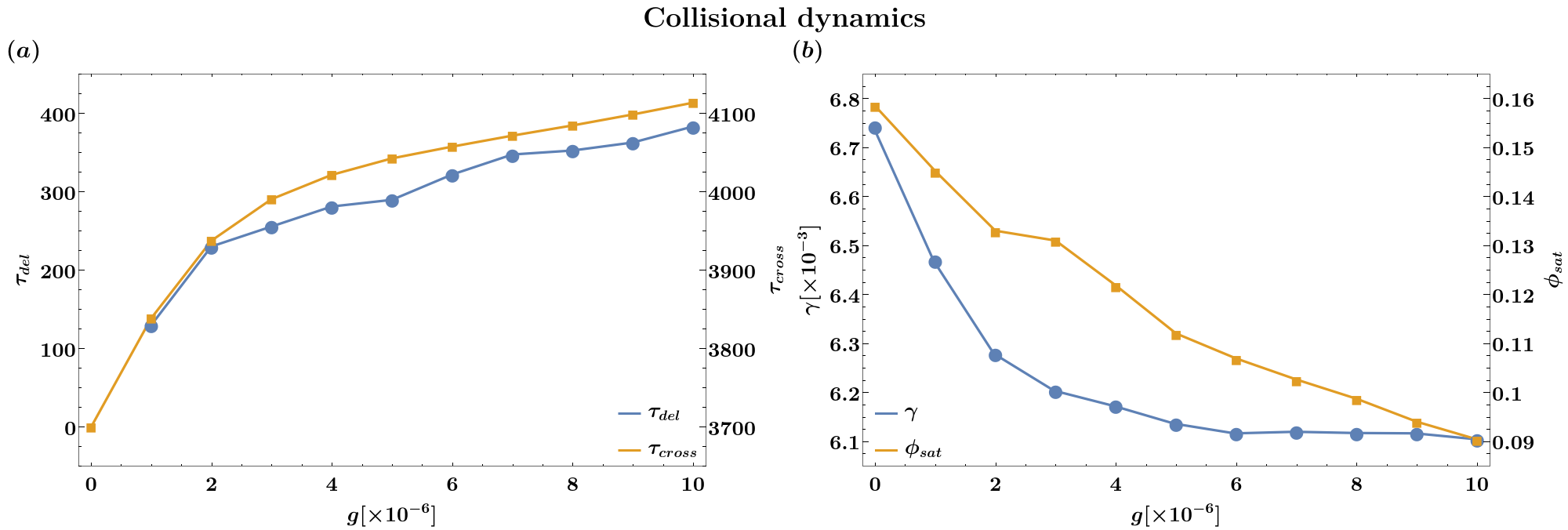}
        \caption{Panel (a): delay time $\tau_{del}$ (blue curve) and crossing time $\tau_{cross}$ (orange curve) as functions of the collisional parameter $g$. Panel (b): growth rate $\gamma$ (blue curve) and saturation amplitude $\phi_{sat}$ (orange curve) as functions of $g$.}
        \label{fig4}
    \end{center}
\end{figure*}

Further analysis was carried out to investigate how some characteristic quantities associated with the instability, represented in Fig.~\ref{fig4}, depend on the collisional term. For instance, the development of the instability can be characterized, together with $\tau_{cross}$, by the delay time $\tau_{del}$, i.e., the additional time required for $|\widehat{\phi}_{1}|$ to reach, in the midpoint of the exponential phase, the same amplitude as in the collisionless case. Therefore, we have repeated the simulation described in Sec.~\ref{sec:results} by varying $g$ from $10^{-6}$ to $10^{-5}$ and measuring these typical times. Their trend as functions of the collisionality is displayed in panel (a): both $\tau_{del}$ (blue curve) and $\tau_{cross}$ (orange curve) show a sharp increase for $g < 3 \times 10^{-6}$, then the growth gradually slows down, showing a slower and more gradual increase for $g > 4 \times 10^{-6}$. Similarly, panel (b) shows the trend of the growth rate $\gamma$ (blue curve) and the saturation amplitude $\phi_{sat}$ associated with $\widehat{\phi}_1$ (orange curve). They exhibit a slight decrease with increasing $g$, while remaining of the same order of magnitude within the margin of numerical errors. 
This highlights that the growth rate of the instability is not strongly affected by the presence of collisions. This result is also consistent with the results discussed in the \hyperref[appendice]{Appendix}. Similarly, the saturation amplitude of the first Fourier mode does not significantly depend on collisional effects.

\subsection{Hermite spectra and enstrophy cascade}
As previously discussed, the Hermite transform of the PDF associated with the evolution of the unstable BGK mode proves to be a powerful diagnostic tool for investigating the formation of small-scale structures in velocity space, in both collisionless and weakly collisional regimes. Moreover, this technique provides an efficient framework for analyzing the development of the turbulent enstrophy cascade in velocity space.

\begin{figure*}[ht]
   \begin{center}
        \includegraphics[]{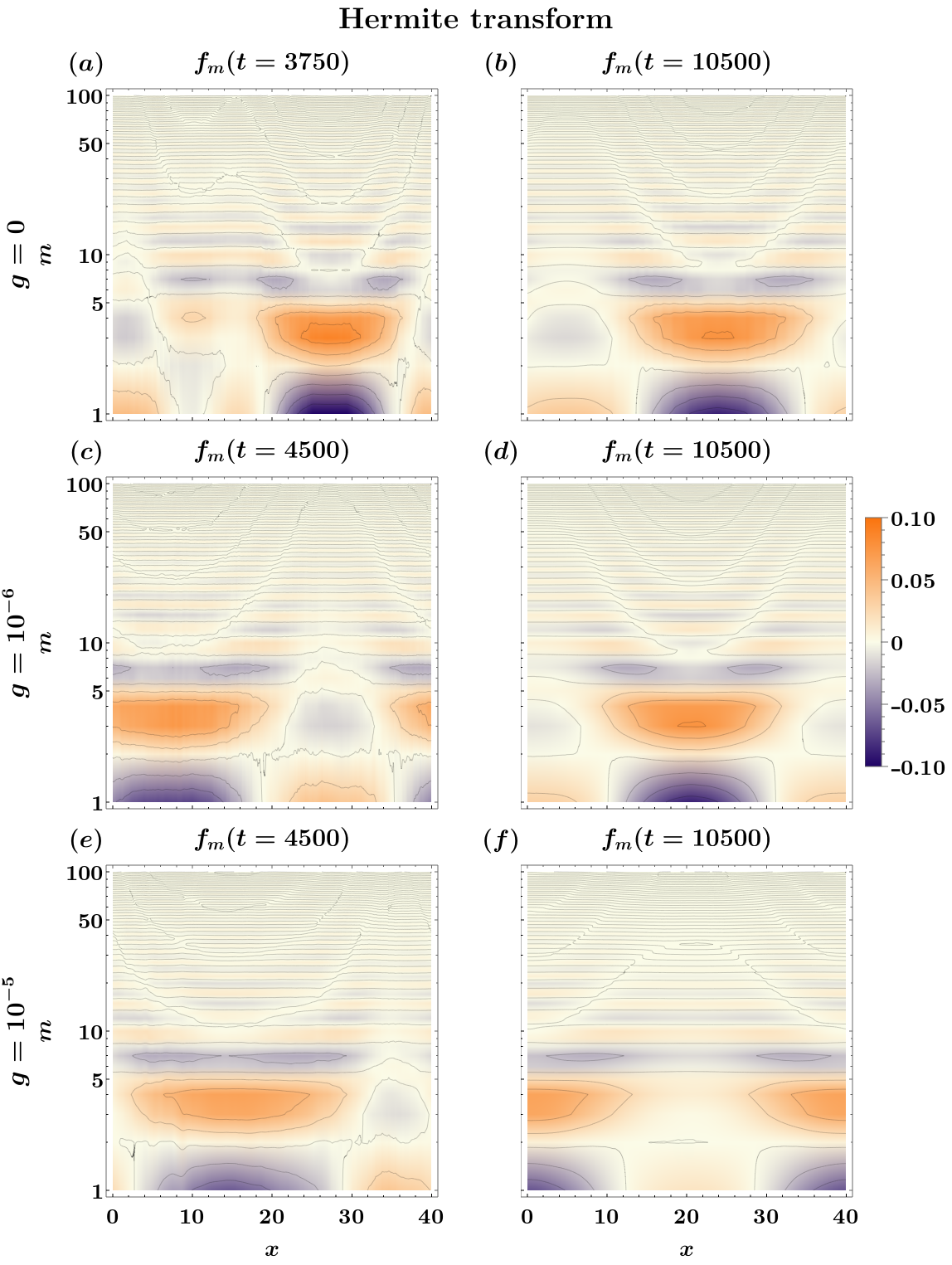}
        \caption{Contour plots of Hermite transform during the merging phase and at the end of the simulation for $g=0$ [panels (a)-(b)], $g=10^{-6}$ [panels (c)-(d)], and $g=10^{-5}$ [panels (e)-(f)].}
        \label{fig5}
    \end{center}
\end{figure*}

Fig.~\ref{fig5} shows the Hermite spectrum associated with the distribution function for $g=0$, [panels (a)-(b)], $g=10^{-6}$ [panels (c)-(d)], and $g=10^{-5}$ [panels (e)-(f)]. It is evaluated immediately after the merging phase ($t=3750$ for $g=0$, $t=4500$ otherwise) and at the final stage of each run ($t=10500$). As expected from the initial, turbulent equilibrium, in each simulation enstrophy has spread throughout the available Hermite modes when the mode $n=2$ is still dominant, therefore small scales are already excited because of the merging of the vortices in the spatial region. We can notice how the spectrum is highly dependent on $x$, with regions particularly rich in enstrophy at low $m$'s that identify the position of the merged structure. On the other hand, in the $m$ direction we observe alternating regions of higher and lower enstrophy with a characteristic pattern that is basically independent of the spatial position.

The runs assume similar behaviors independently of the collisional frequency. However, for $t=t_{max}$, the distribution function associated with $g=10^{-5}$ exhibits less pronounced peaks due to the dissipative effects introduced by the Dougherty operator. Indeed, even though this case can be defined as weakly collisional, because the nonlinear dynamics is much quicker than the collisional time $\nu_0^{-1}$, the latter is comparable with $t_{max}$, so we expect that a broad range of scales is affected by collisions. If $g=10^{-6}$, instead, $t_{max} \ll \nu_0^{-1}$, therefore the dissipation would only involve the highest available Hermite modes.

\begin{figure*}
   \begin{center}
        \includegraphics[]{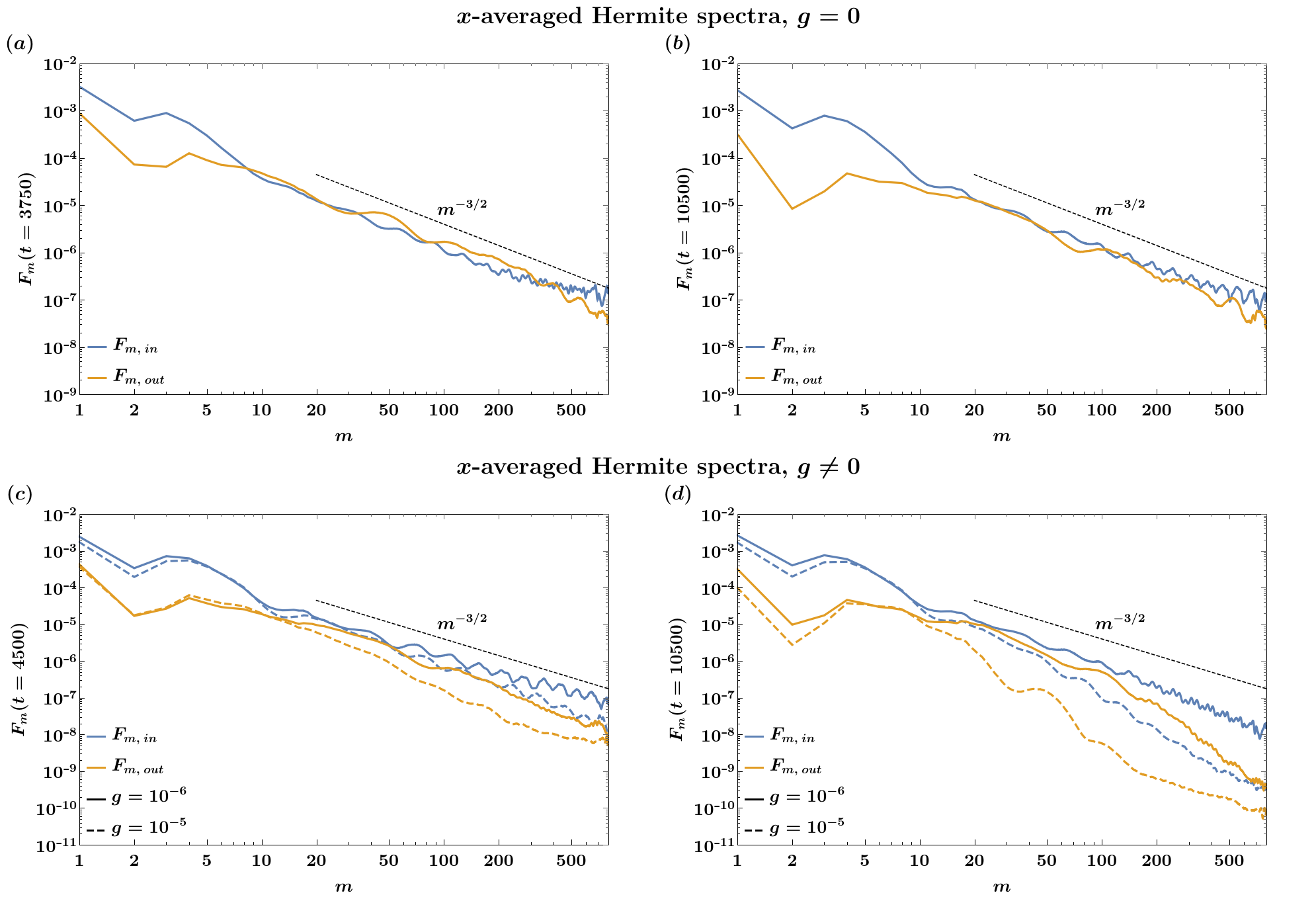}
        \caption{Spatially-averaged Hermite spectrum during the merging phase and at the end of the simulation in the collisionless regime [panels (a)-(b)] and the weakly collisional regime [panels (c)-(d)], for $g=10^{-6}$ (solid lines) and $g=10^{-5}$ (dashed lines). Blue and orange curves represent the enstrophy content inside and outside the vortex, respectively, while the black dashed line represents the expected trend $m^{-3/2}$ for electrostatic turbulence.}
        \label{fig6}
    \end{center}
\end{figure*}

To quantitatively analyze how enstrophy is distributed across velocity space, we define the $x$-averaged Hermite spectrum as:
\begin{equation}
    F_{m}(t) = \langle f_{m}^{2} \rangle_{x} = \frac{1}{L_{x}} \int_{0}^{L_{x}} f_{m}^{2}(x,t) \, dx .
\end{equation}
This approach does not allow to separate the contributions of trapped and untrapped particles into the development of the cascading mechanism. To do so, denoting by $\Omega$ the $x$ region occupied by the vortex at each time, we define the internal and external spectrum as:
\begin{gather}
F_{m,\; in}(t) = \frac{1}{L_{x}} \int_{x \in \Omega} f_{m}^{2}(x,t) \, dx, \label{Fin} \\[5pt]
F_{m,\; out}(t) = \frac{1}{L_{x}} \int_{x \notin \Omega} f_{m}^{2}(x,t) \, dx, \label{Fout}
\end{gather}
constructed in such a way that $F_m(t) = F_{m,\; in}(t) + F_{m,\; out}(t)$. Applying Eqs.~\eqref{Fin}-\eqref{Fout} to the spectra depicted in Fig.~\ref{fig5}, we obtain the averages shown in Fig.~\ref{fig6} as functions of $m$. More specifically, $F_{m,\; in}$ represents the enstrophy contribution inside the vortex, while $F_{m,\; out}$ represents the contribution outside the vortex.

Starting from the collisionless case [panels (a)-(b)], we observe how, at large scales (low $m$'s), enstrophy is predominantly concentrated within the vortex. As we move toward smaller scales (high $m$'s), the enstrophy amount inside and outside the vortex becomes comparable. On the other hand, when collisions are turned on, enstrophy dissipation seems to be more effective outside the vortex rather than inside. This effect can be already appreciated during the merging process [panel (c)]: as expected, at high $m$’s both $F_{m, \; in}$  and $F_{m, \; out}$ decrease with $g$, but when the Dougherty operator is included, \(F_{m,\; out} / F_{m,\; in} < 1\) regardless of the collisional frequency. This is more evident at late times [panel (d)]: $F_{m, \; out}$ is around one order of magnitude less than $F_{m, \; in}$ for $m \gtrsim 100$ when $g=10^{-6}$ (solid lines) and for $m \gtrsim 20$ when $g=10^{-5}$ (dashed lines).

This enhancement of enstrophy dissipation outside the vortex could be explained by the different profile of the PDF in the two regions. Without collisions, the high grid resolution allows structures to survive throughout the whole Hermite range even after thousands of plasma periods, as inferred from panels (a)-(b). On the other hand, when the Dougherty operator is added, we expect that the enstrophy flux ceases to be conserved, and therefore the Hermite spectrum of the PDF displays a cutoff, after a critical value $m^\star$, similarly to the Kolmogorov model for turbulence in physical space. As discussed in Ref.~\onlinecite{pezzi2019fourier}, $m^\star$ for a fixed value of collisional frequency increases with the Fourier amplitude of the electric field, and consequently with the enstrophy amount at large scales. Therefore, the enhanced deviation from the Maxwellian in the region inside the vortex justifies the better efficiency of the enstrophy cascade associated with $F_{m, \; in}$ rather than  $F_{m, \; out}$.

\begin{figure*}
   \begin{center}
        \includegraphics[]{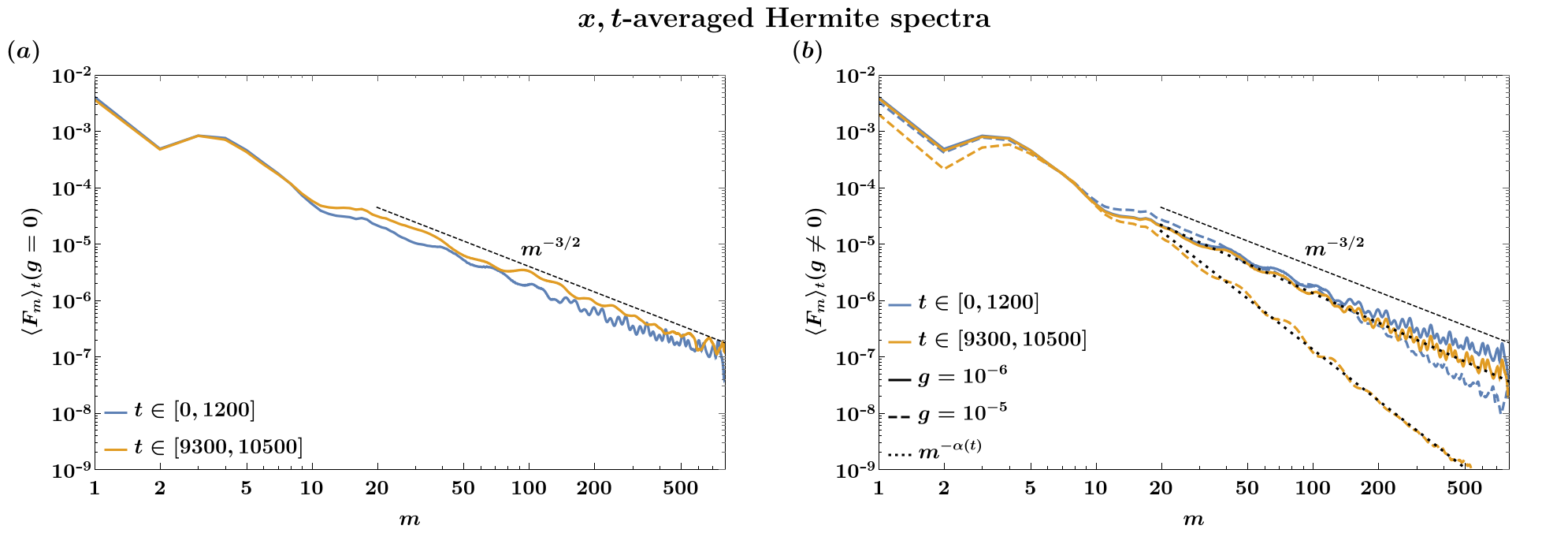}
        \caption{Spatially and temporally averaged Hermite spectrum at early times (blue curves) and final times (orange curves) for the collisionless regime [panel (a)] and the weakly collisional regime [panel (b)], for $g=10^{-6}$ (solid lines) and $g=10^{-5}$ (dashed lines). The black dashed line represents the expected trend $m^{-3/2}$ for electrostatic turbulence, while the black dotted lines indicate the best fits $m^{-\alpha}$ associated with intermediate scales.}
        \label{fig7}
    \end{center}
\end{figure*}

We now consider the full spectrum \(F_{m}\) and investigate its trend. According to the model discussed in Refs.~\onlinecite{servidio2017magnetospheric,pezzi2019fourier}, $F_{m} \sim m^{-3/2}$ for $m \gg 1$ when the nonlinear term associated with the electric field dominates over the linear advection term and the cascade flux is kept constant. In order to get a clear trend to compare with theory, we calculate the mean spectrum not only in $x$ but also in $t$, by considering time windows with fixed amplitude at different stages of the system. Fig.~\ref{fig7} shows the $x,t$-averaged spectra $\langle F_{m} \rangle_t$, calculated for $t \in [0, 1200]$ (blue curves) and $t \in [9300, 10500]$ (orange curves). Panels (a) and (b) refer to the collisionless and collisional cases, respectively.
 
Specifically, when $g=0$ the spectrum closely follows the expected power law (black dashed line). A numerical fit of the form \(m^{-\alpha}\), performed in the range \(m \in [20, 500]\), yields $\alpha \simeq 1.50$ for $t<1200$ and $\alpha \simeq 1.52$ for $9300<t<10500$, showing that the high grid resolution maintains a stable, undissipated cascade throughout the whole Hermite space. When $g \neq 0$, enstrophy is gradually damped at small scales, but the collisionality is still too low for the Dougherty operator to dominate the system dynamics, and therefore no exponential decay is observed. Instead, a power law with an increasing slope is identified: numerical fits, represented by dotted lines in panel (b), quickly deviate from the collisionless case. More specifically, when $g=10^{-5}$ (orange dashed lines), \(\alpha \simeq 1.73\) at early times and \(\alpha \simeq 3.01\) at later times.

\begin{figure*}[b!]
\begin{center}
        \includegraphics[]{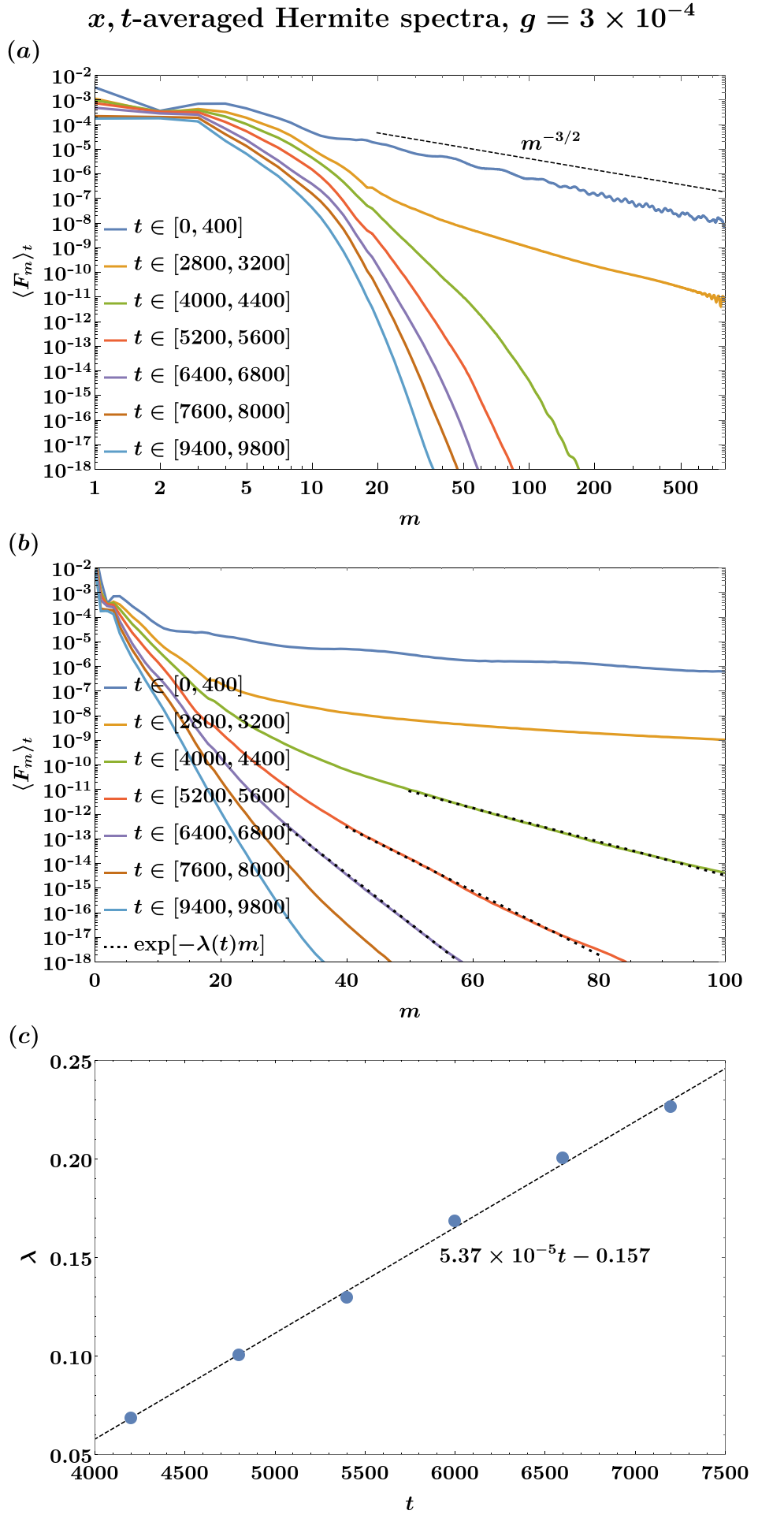}
        \caption{Panels (a)-(b): spatially and temporally averaged Hermite spectrum over different time intervals for $g = 3 \times 10^{-4}$ in log-log scale [panel (a)] and semi-log scale [panel (b)]. The black dashed line represents the expected trend $m^{-3/2}$ for electrostatic turbulence, while the black dotted lines indicate the best fits $e^{-\lambda m}$ associated with intermediate scales. Panel (c): linear fit of the decay parameter $\lambda(t)=\alpha_0t+\beta_0$.}
        \label{fig8}
    \end{center}
\end{figure*}

In order to check the features of enstrophy dissipation in this scenario and whether it is in agreement with Ref.~\onlinecite{pezzi2019fourier}, we performed a simulation with a higher collisionality, setting \( g = 3 \times 10^{-4} \), whose mean spectrum is depicted in Fig.~\ref{fig8}. Due to the faster evolution in this regime, $F_{m}$ is averaged over a shorter temporal window. Panel (a) shows that, at early times, the spectrum exhibits a clear power law behavior similar to what is observed for $g=10^{-5}$. However, after waiting for enough time, the onset of the expected spectral cutoff becomes evident, in agreement with the behavior of the phase space distribution (not shown here) for this choice of $g$. The dissipation is sufficient to contrast the instability and hole structures disappear around $t=3000$, i.e., at the formation of the spectral cutoff. Consequently, we expect that after this time, at least at small scales, the Dougherty operator dominates over the linear and nonlinear advection terms of the Vlasov equation.

If this hypothesis holds, then the decay trend can be mathematically justified by adopting the model introduced by Pezzi et al.~\cite{pezzi2019fourier}, which expresses the collisional term in the Fourier-Hermite space as a series of creation and annihilation operators to apply to the transformed PDF. Considering that, for \(m \gg 1\), each application of a creation or annihilation operator introduces a factor of \(\sqrt{m}\), the Dougherty operator in the Hermite space can be approximately expressed as:
\begin{equation}
    C_{m} \sim -\nu_{0} m F_{m},
    \label{D}
\end{equation}
where $\nu_{0}=-g\ln{g}/(24\pi) \simeq 3.23 \times 10^{-5}$. If $C_{m}$ dominates over the other terms of the Hermite-transformed Vlasov equation, then it becomes:
\begin{equation}
    \frac{\partial F_{m}}{\partial t} \sim -\nu_{0} m F_{m},
\end{equation}
whose solution takes the exponential form:
\begin{equation}
    F_{m}(t) \sim F_{m}(t_{0}) \, e^{-\nu_{0} m (t - t_{0})}.
\end{equation}

The spectrum cutoff is in agreement with this trend, as we can observe in Fig.~\ref{fig8} (b), representing $\langle F_{m} \rangle_t$ on a semi-log scale. Indeed, in the time range $4000<t<7000$, the spectrum can be fitted with the exponential law $e^{-\lambda(t) m}$ (black dotted lines).
The parameter $\lambda$ tends to follow a linear trend $\lambda(t)=\alpha_0t+\beta_0$, with a fitted slope \(\alpha_{0} \simeq 5.37 \times 10^{-5}\) very close to $\nu_{0}$, as expected [panel (c)]. Moreover, \(\beta_{0} \simeq -0.157\), giving an estimate of the parameter $t_0$ in Eq.~\eqref{D} as \(- \beta_{0}/\alpha_{0} \simeq 2920\), i.e., the moment at which the Dougherty operator begins to dominate. This is indeed compatible with the observed instant of the simulation when the enstrophy trend switches from a power law to an exponential decay ($t \simeq 3000$). This numerically validates the theoretical framework proposed by Pezzi et al.~\cite{pezzi2019fourier} when collisional effects become significant.

\section{Summary}\label{sec:summary}
In this paper, we studied the dynamics of an unstable one-dimensional BGK mode through high-resolution 1D-1V Vlasov-Poisson numerical simulations with kinetic electrons and motionless ions. The system was studied in both collisionless and weakly collisional cases. We perturbed an initial BGK equilibrium with a large-scale density noise, exciting an electrostatic instability. This instability caused the merging of two initial phase-space vortices into one coherent hole. We studied the phase-space dynamics of the electron PDF relative to electrons by turning on and off the collisional term, represented by the Dougherty operator. Our findings indicated that collisions tend to delay the onset of the instability and, as a consequence, the phase-space merging process.

To further confirm the presence of this delay, we analyzed the temporal evolution of Fourier modes $n = 1$ and $n = 2$. In both regimes, the $n = 1$ mode exhibited exponential growth of the form $e^{\gamma t}$, with the growth rate $\gamma$ remaining of the same order of magnitude. The key difference, which highlights the effect of weak collisions, lies in the crossing time between the $n = 1$ and $n = 2$ modes, which was delayed in the weakly collisional case. In line with this, we further investigated how specific quantities vary with the collisional parameter. We found that, within numerical accuracy, the growth rate of the instability and the saturation amplitude of the $n = 1$ mode remained unchanged. In contrast, both the delay time and the crossing time were significantly affected, confirming the phase-space delay observed earlier.

To analyze the formation of small-scale structures not only in physical space, but also in velocity space, we employed the Hermite transform of the electron distribution function. This tool, implemented in our code, enabled us to examine how collisions influence the enstrophy cascade in the system.
From our analysis, we observed that once the instability developed, a vortex merging process occurred in physical space, while in Hermite space an enstrophy cascade emerged and persisted until the end of the simulation in both the collisionless and weakly collisional regimes. To investigate how the enstrophy cascade is influenced by the spatial position, we analyzed the $x$-averaged Hermite spectrum inside and outside the vortex. In the collisionless case, the enstrophy contribution remained comparable in both regions. In contrast, collisions caused enstrophy at small scales to be more abundant within the vortex. This implies a stronger dissipation of enstrophy outside the vortex region. 

A more quantitative investigation was achieved by comparing the average Hermite spectrum with the characteristic power-law scaling of electrostatic turbulence $m^{-3/2}$, in agreement with Ref.~\onlinecite{servidio2017magnetospheric}. This allowed us to verify that, in the collisionless case, enstrophy cascades followed the theoretical expectations, with a flux kept constant both at early and late times. In the presence of very weak collisions ($g=10^{-5}$), small-scale enstrophy was dissipated, but a power-law scaling was still maintained throughout the simulation. On the other hand, when the collisional rate was higher ($g = 3 \times 10^{-4}$), the Hermite spectrum exhibited a clear cutoff, with an exponential decay at large $m$'s, suggesting that in this regime the Dougherty operator becomes dominant in the Vlasov equation.

\begin{acknowledgments}
This project has received funding from the European Union’s Horizon Europe research and innovation programme under Grant Agreement No. 101082633 (ASAP) and from the ASI project “Attività di Fase A per la missione Plasma Observatory” (2024-15-HH.0). F.V. acknowledges the support of the PRIN 2022 project “The ULtimate fate of TuRbulence from space to laboratory plAsmas (ULTRA)” (2022KL38BK, Master CUP: B53D23004850006), funded by the Italian Ministry of University and Research. The numerical simulations presented in this paper were performed on the ALARICO cluster at the University of Calabria. This study was carried out within the Space It Up project funded by the Italian Space Agency, ASI, and the Ministry of University and Research, MUR, under Contract No. 2024-5-E.0-CUP No. I53D24000060005. 
\end{acknowledgments}

\section*{Author declarations}

\subsection*{Conflict of interest}

The authors have no conflicts to disclose.

\section*{Data availability}

The data that support the findings of this study are available from the corresponding author upon reasonable request.

\appendix
\section*{Appendix: Growth rate estimation for inhomogeneous equilibria}
\label{appendice}
\renewcommand{\theequation}{A\arabic{equation}}
\setcounter{equation}{0} 
In this appendix, we discuss in detail an approximate method to predict the instability growth rate associated with spatially-dependent BGK equilibria. We apply this technique to the numerical PDF associated with the propagation of EAWs analyzed in this paper and an inhomogeneous two-hole distribution built to satisfy the time-independent Vlasov equation with a very small error, quantifying the accuracy of the predicted growth/decay rates with respect to the observed behavior of perturbations. 
\subsection*{A.1\hspace{1em}Approximated linear dispersion relation} \label{appA}
Following Ref.~\onlinecite{manfredi2000stability}, we consider a generic equilibrium PDF $F(x,v)$, whose associated field is $E_0(x)$, which is perturbed by a first-order fluctuation $\delta f (x,v,t)$ that corresponds to an electric perturbation $\delta E(x,t)$. Substituting $f=F+\delta f$ and $E=E_0 + \delta E$ into Eq.~\eqref{eqn:f} leads to:
\begin{equation} \label{linear}
\frac{\partial \delta f}{\partial t} + v\frac{\partial \delta f}{\partial x} - \delta E \frac{\partial F}{\partial v} - E_0 \frac{\partial \delta f}{\partial v} = 0, 
\end{equation}
where higher-order terms have been neglected. If $E_0 \ll 1$, then the last term is much smaller than the third and can be ignored.

The previous equation can be directly manipulated by following the Landau theory only when $F$ is not dependent on $x$. However, we can, with some approximation, obtain a homogeneous version of Eq.~\eqref{linear}. If we treat single-scale fluctuations $\delta f = \delta \hat{f} e^{i k x}$ and $\delta E = \delta \hat{E} e^{i k x}$, where $k$ is a much lower wavenumber than those associated with $F$ oscillations, then we can perform the average $\langle ... \rangle =  \int_{\bar{x}-\delta x}^{\bar{x}+\delta x} ... dx / \left( 2\delta x \right)$ in a window where $\delta E$ varies much slower than $\partial F / \partial v$, in such a way that $\langle \delta E \left( \partial F / \partial v  \right) \rangle \approx \langle \delta E \rangle d \langle F \rangle / d v$:
\begin{equation} 
\left( \frac{\partial \delta \hat{f}}{\partial t} + ikv \delta \hat{f} - \delta \hat{E} \frac{d \langle F \rangle}{d v} \right)  \langle e^{ikx} \rangle\approx 0 .
\end{equation}
Assuming that in the selected window the average oscillation is nonzero, then the previous equation can be Laplace-transformed, leading to the usual Landau treatment of plasma waves based on the searching for the roots of the dielectric function $D=D_R+iD_I$ associated with $f_0(v)=\langle F (x,v) \rangle$. Following Ref.~\onlinecite{valentini2012undamped}, the frequency of the perturbation $\omega_R + i \omega_I$, provided that $\omega_I \ll \omega_R$, can now be approximately determined by the conditions:
\begin{gather}
D_R(k,\omega_R) = 1 - \frac{1}{k^2} \mathcal{P} \int_{-\infty}^{+\infty} \frac{f_0'(v)}{v - \omega_R / k} \, dv = 0, \label{omegaR} \\[5pt]
\omega_I (k) = -\frac{D_I (k,\omega_R)}{\partial_\omega D_R \bigr\vert_{k,\omega_R}} =  \frac{\pi}{k^2}\frac{f_0'\bigr\vert_{\omega_R/k}}{\partial_\omega D_R \bigr\vert_{k,\omega_R}}, \label{omegaI}
\end{gather}
where $\mathcal{P}$ denotes the principal value of the integral and $\partial_\omega D_R \bigr\vert_{k,\omega_R}$ is the partial derivative of $D_R$ evaluated at $(k,\omega_R)$.

This technique can be useful for deducing how fluctuations behave in the region of trapped particles and how the roughness of the plateau can influence their growth rate. For this reason, for the analyzed two-hole equilibria, perturbed by density fluctuations at $k=k_0$, we choose $\bar{x}$ as the center of one of the two vortices. We build $F$ on an equilibrium potential $\phi_0 \propto \cos(k_2 x -\varphi(t))$, in such a way that the width of the plateau, i.e., $\delta v \propto \sqrt{1-\cos(k_2x-\varphi)}$, reaches its maximum $\delta v_{max}$ for $x=\bar{x}(t)$. With this PDF, choosing $\delta x= L_x/6$ allows us to define $f_0$ as a mean of velocity distributions whose plateau is at least $\delta v_{max}/2$ wide, in such a way as to capture the behavior of resonant particles without using a spatial average over an excessively large interval.

\subsection*{A.2\hspace{1em}Construction of analytical two-hole equilibrium}
In this section, we construct a two-hole BGK state which, unlike quasi-stationary EAW turbulence, is characterized by perfectly flat plateaus in the resonant region, to numerically observe how this feature influences the transition of trapped particles from one hole to the other when some large-scale noise is added, compared with the case mainly studied in this work. We use a method similar to that described by Demeio and Holloway \cite{demeio1991numerical}: starting from a given potential shape $\phi_0(x)$, we define an equilibrium PDF $g(x,v)$ with a plateau centered at $v=v_p$ such that $(g(x-v_pt,v-v_p),\phi_0(x-v_pt))$ approximately solves the Vlasov-Poisson system. To do so, we build $g$ as a function of the energy of particles $\mathcal{E}(x,v)=(v-v_p)^2/2-\phi_0(x)$ only, with some refinements with respect to the aforementioned work to guarantee better consistency with the assigned potential.

We set $\phi_0=-\epsilon \cos(k_2x)$: in this way, the equation of curves with energy $\mathcal{E}$ and the separatrices (such that $\mathcal{E} = \epsilon$) are respectively given by:
\begin{gather}
v_{\mathcal{E}}^{\pm}(x)=v_p \pm \sqrt{2[\mathcal{E}-\epsilon \cos(k_2x) ]}, \\[5pt]
v^{\pm}(x)=v_p \pm \sqrt{2 \epsilon[1 - \cos(k_2x) ]} .
\end{gather}
Defining $v_{\mathcal{E},0}^{\pm}(x,v) = v_p \pm \sqrt{2(\mathcal{E} (x,v)-\epsilon )}$, we then set:
\begin{equation}
g_{\beta}(x,v) = \begin{cases}
f_\beta (v) & \text{if}\;x=0,L \\
f_\beta (v_p) & \text{if}\;v_-(x) \leq v \leq v_+(x) \\
f_\beta (v_{\mathcal{E},0}^{-}(x,v)) & \text{if}\; v \leq v_-(x) \\
f_\beta (v_{\mathcal{E},0}^{+}(x,v)) & \text{if}\; v \geq v_+(x)
\end{cases},
\end{equation}
where $f_\beta (v) = C_\beta e^{-v^2/[2(1-\beta)]}$. This corresponds to a generalization of the method presented in Ref.~\onlinecite{demeio1991numerical}, corresponding to $\beta=0$. At the same time, we expect that $g_\beta$ satisfies the stationary Vlasov equation, assuming $\phi_0$ as the electric potential, with $\beta$ that can be tuned such that the solution of the Poisson equation:
\begin{equation}
\frac{d^2\phi_\beta}{dx^2} = n_\beta - 1, 
\end{equation}
where $n_\beta=\int_{-\infty}^\infty g_\beta \, dv$, is as close as possible to $\phi_0$.

\begin{figure*}
   \begin{center}
        \includegraphics[]{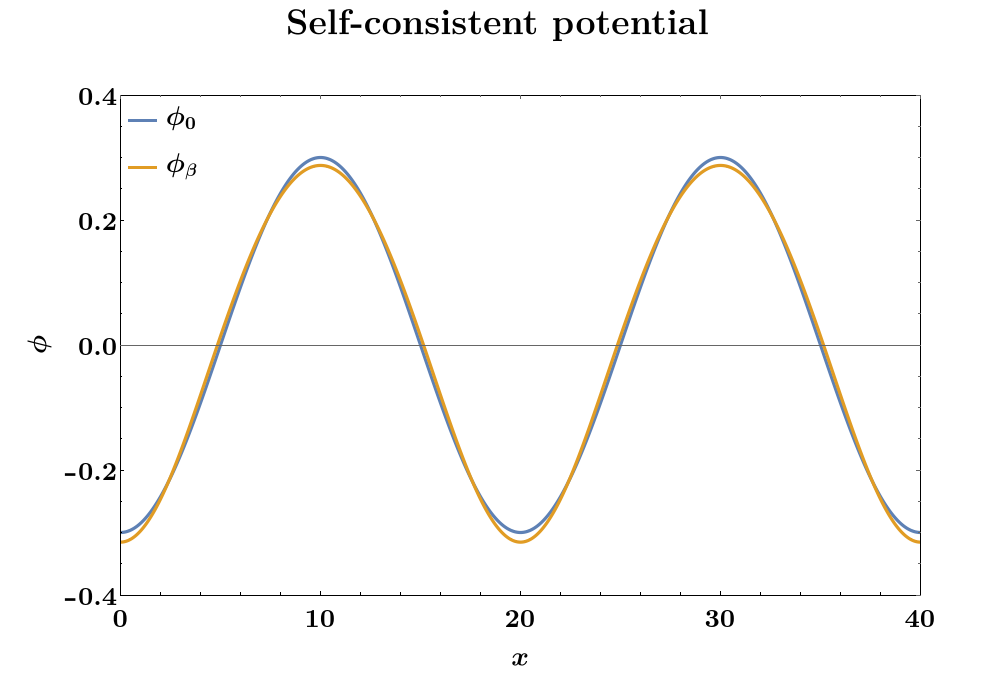}
        \caption{Comparison between $\phi_0$ (blue curve) and $\phi_\beta$ (orange curve) for $\epsilon =0.3$, $v_p=1.7$, and $\beta=0.12804$.}
        \label{potential}
    \end{center}
\end{figure*}

It is possible to show that:
\begin{equation}
n_\beta (x) = C_\beta \left\lbrace \int_{-\infty}^\infty \frac{\left| v-v_p \right| e^{-v^2/[2(1-\beta)]}}{\sqrt{(v-v_p)^2+2\epsilon[1-\cos(k_2x)]}} \, dv + 2\sqrt{2\epsilon[1-\cos(k_2x)]} e^{-v_p^2/[2(1-\beta)]}\right\rbrace .
\end{equation}
Once $\beta$ is fixed, the normalization factor $C_\beta$ can be set in such a way that $n_\beta$ is 1 on average, so that $\langle \phi_\beta \rangle_x = \langle \phi_0 \rangle_x = 0$. Observing the previous equation, one can find that the odd Fourier components of $n_\beta$ are always null, while $\beta$, for some ranges of $v_p$ and $\epsilon$, can be set so that the $k=k_2$ component coincides with $-\epsilon k_2^2$. This guarantees that $\phi_\beta$ and $\phi_0$ harmonics differ only for even modes with $k \geq k_4$. More specifically, setting $\epsilon=0.3$ and $v_p=1.7$ in the same physical domain of EAW simulations ($L_x=40$), this value is $\beta=0.12804$, leading to $\phi_\beta\sim\phi_0$ (see Fig.~\ref{potential}), confirmed by the stationarity of the numerical runs where $f$ is initially set equal to $g_\beta$.

\subsection*{A.3\hspace{1em}Numerical results}

\begin{figure*}
   \begin{center}
        \includegraphics[]{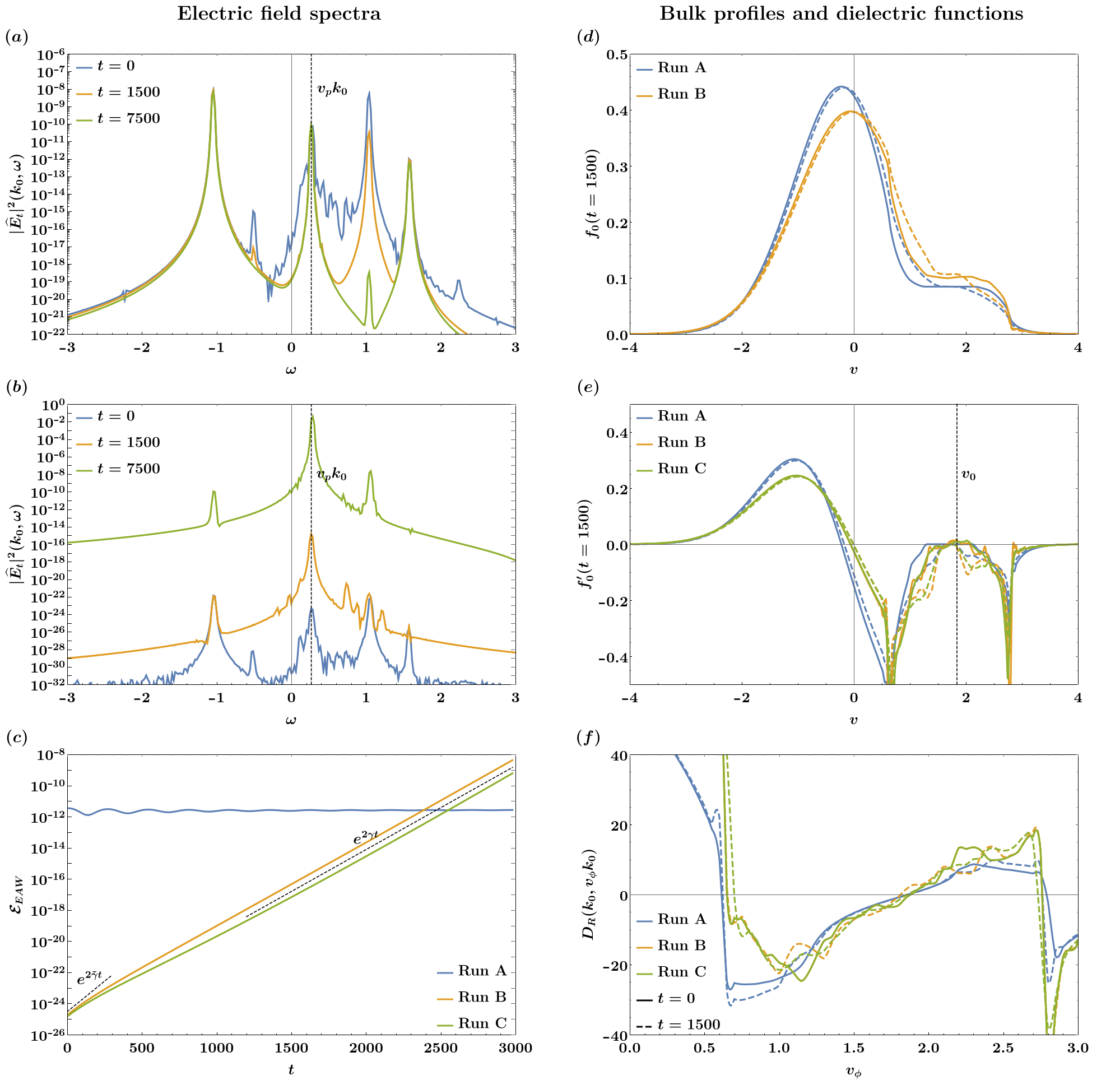}
        \caption{Panels (a)-(b): electric field spectrum evaluated at $k_0$ as a function of $\omega$ at several times, for Run A [panel (a)] and Run B [panel (b)], compared with frequency $v_p k_0$ (black dashed line). Panel (c): time evolution of electric energy associated with the EAW mode, compared with growth rates $\gamma$ and $\tilde{\gamma}$ (black dashed lines). Panels (d)-(e): $x$-averaged PDF $f_0$ [panel (d)] and its derivative $f_0'$ [panel (e)]. Means over $[\bar{x}-L_x/6, \bar{x}+L_x/6]$ (solid lines) and the whole spatial domain (dashed lines) are compared. Panel (f): real part of the dielectric function $D_R$ associated with fixed wavenumber $k_0$ and varying phase velocity $v_\phi$, at $t=0$ (solid lines) and $t=1500$ (dashed lines).}
        \label{figapp}
    \end{center}
\end{figure*}

We focus on three numerical runs. Run A describes the collisionless dynamics of a PDF whose equilibrium at $t=0$ is the distribution $g_\beta(x,v)$ described in the previous paragraph. In this case, the starting density perturbation is $\delta n(x,t=0)=A\sin\left( k_0x \right)$, with $A=7.4022 \times 10^{-6}$. Run B and Run C, instead, represent the same scenarios of Sec.~\ref{sec:numericalmodel} with $g=0$ and $g=10^{-5}$, respectively, but with reduced grid resolution ($N_x \times N_v = 2048 \times 2001$). The numerical results are summarized in Fig.~\ref{figapp}.

First, we focus on the time evolution of the 2D Fourier spectrum of the electric field, given by:
\begin{equation}
\widehat{E}_t \left( k, \omega \right) = \frac{1}{L_xT} \int_{t}^{t+T} \int_{0}^{L_x} E (x,t') e^{-i\left( kx -\omega t' \right)} \, dx \, dt' ,
\end{equation}
observing the behavior of the cut $k=k_0$ at several times ($T =1024 \Delta t$). In panel (a), it is possible to observe how the initial perturbation mainly excites a pair of Langmuir waves, but also, as expected, an EAW whose phase velocity is highly compatible with $v_p$. Even some higher-order modes are present. As time passes, the backward Langmuir wave and the EAW mode maintain an approximately constant amplitude, while the forward Langmuir wave is progressively damped. The leftward wave stores most of the electric energy throughout the whole simulation, so the fundamental mode results in being marginally stable in the simulation time. On the other hand, Run B spectrum [panel (b)] shows that the EAW effectively grows thanks to the dynamics of the trapping particles, dominating the Langmuir oscillations. Run C presents qualitatively similar features.

The energy associated with the EAW is estimated as $\mathcal{E}_{EAW} (t) = \sum_{\omega \approx v_pk_0} \vert \widehat{E}_t \left( k_0, \omega \right) \vert^2$. As shown in panel (c), the mode is marginally stable in Run A and unstable in Runs B-C, with a growth rate $\gamma \simeq 6 \times 10^{-3}$ regardless of $g$. This rate is compatible with values observed in Sec.~\ref{sec:results}, even though slightly smaller because of the lower grid resolution. In addition, the isolation of the mode even when it is not dominant allows us to observe a transient rate $\tilde{\gamma}\simeq 9 \times 10^{-3}$ at very early times.

Such a different behavior between Run A and Runs B-C is substantially due, as discussed before, to the derivative of the bulk PDF in the resonant zone, which is zero by construction in Run A, leading to a null growth rate according to Eq.~\eqref{omegaI}. The importance of $f_0$ shape at resonance is confirmed by observing its plots represented in panel (d). The assumed profile for $f_0$ is given by averaging the full PDF as described in Sec.~\hyperref[appA]{A.1} (solid lines). This allows us to note that the plateau in Run B (and Run C, not shown here), is slightly rough despite the $x$ average, while in Run A it is obviously flat. The derivative $f_0'$ is shown in panel (e). Run B (orange line) and Run C (green line) at $t=1500$ are slightly different, especially in the plateau region. However, unlike $\langle f \rangle_x$ (dashed lines), both assume small, positive values in a narrow velocity range, not excluding positive growth rates as actually observed.


We calculate $D_R$ as expressed in Eq.~\eqref{omegaR} for $k=k_0$, as a function of the phase velocity $v_\phi \equiv \omega_R / k_0$, in seven cases, i.e., for each run at $t=0$ and $t=1500$, and for Run A at $t=7500$ (because the linearity of the system still holds). Being the examined $f_0$ profiles comparable at large scales, the dielectric function trend in panel (f) is similar in the treated cases. More specifically, we can find three roots of $D_R$, in agreement with analogous analysis by Valentini et al. \cite{valentini2012undamped}. Two of these are at the corner of the velocity plateau, corresponding to damped modes. The third is inside the plateau and corresponds to the value $v_0$ to which we are interested. $v_0$ varies from 1.82 to 1.88, so it assumes values larger than $v_p=1.7$. This can happen for EAW modes that do not exactly fall on the thumb curve defined by the roots of $D_R$ in the infinitesimal plateau limit (corner modes, see Ref.~\onlinecite{valentini2012undamped}). 

In Tab.~\ref{tabella}, we report the observed values of $f_0'\bigr\vert_{v_0}$ and $\partial_\omega D_R \bigr\vert_{k_0, \omega_0 }$, with $\omega_0 \equiv v_0 k_0$: this gives, following Eq.~\eqref{omegaI} for the imaginary part of the EAW frequency, an estimate of the growth rate $\gamma_0$:
\begin{equation}
\gamma_0 = \frac{\pi}{k^2_0}\frac{f_0'\bigr\vert_{v_0}}{\partial_\omega D_R \bigr\vert_{k_0,\omega_0}} .
\end{equation}
For the EAW equilibrium, we find a higher value of $\gamma_0$ at $t=0$ than $t=1500$, in agreement with the observation $\gamma < \tilde{\gamma}$. At $t=0$, Run B and Run C start from the same PDF, leading to the same $\gamma_0$ estimate. On the other hand, because of the difference in the collisional term, they evolve into slightly different $f_0$ profiles as time passes, but the growth rate remains similar as expected by observations. Indeed, the relative error $\Delta \gamma$, i.e. $(\gamma_0-\tilde{\gamma})/\tilde{\gamma}$ at $t=0$, $(\gamma_0-\gamma)/\gamma$ at $t=1500$, never exceeds 25$\%$.


\begin{table*}
\caption{Estimated growth rate $\gamma_0$ and its error $\Delta \gamma$, paired with Landau theory parameters, associated with Runs A-C.\label{tabella}}
\begin{ruledtabular}
\begin{tabular}{ccccccc}
 Run & $t$ & $v_0$ & $f_0'\bigr\vert_{v_0}$ & $\partial_\omega D_R \bigr\vert_{k_0, \omega_0}$ & $\gamma_0$ & $\Delta \gamma$  \\
 \hline
 A & 0 & 1.87 & $\simeq 0$ & 13.1 & $\simeq 0$  & /  \\
 A & 1500 & 1.87 & $\simeq 0$ & 12.8 & $\simeq 0$  & /  \\
 A & 7500 & 1.86 & $\simeq 0$ & 12.2 & $\simeq 0$  & / \\
 B & 0 & 1.88 & $1.39 \times 10^{-2}$ & 25.5 & $1.09 \times 10^{-2}$ & 21 $\%$ \\
 B & 1500 & 1.82 &  $1.46 \times 10^{-2}$ & 43.4 & $6.73 \times 10^{-3}$ & 12 $\%$ \\
 C & 0 & 1.88 & $1.39 \times 10^{-2}$ & 25.5 & $1.09 \times 10^{-2}$ & 21 $\%$ \\
 C & 1500 & 1.86 & $8.38 \times 10^{-3}$ & 23.0 & $7.30 \times 10^{-3}$ & 22 $\%$ \\
\end{tabular}
\end{ruledtabular}
\end{table*}

\nocite{*}
\bibliography{biblio_ZANELLI}
\bibliographystyle{apsrev4-1}

\newpage

\end{document}